%% file: main.tex
\documentclass[conference]{IEEEtran}
\IEEEoverridecommandlockouts

\usepackage{pgfplots}
\usepackage{graphicx}
\usepackage{textcomp}
\usepackage{xcolor}
\usepackage[a4paper, total={184mm,239mm}]{geometry}
\def\BibTeX{{\rm B\kern-.05em{\sc i\kern-.025em b}\kern-.08em
    T\kern-.1667em\lower.7ex\hbox{E}\kern-.125emX}}

\usepackage{graphicx}
\usepackage{amsmath,amssymb,amsfonts}
\usepackage[noend,linesnumbered,ruled,vlined]{algorithm2e}
\usepackage{booktabs}
\usepackage{listings}
\usepackage{microtype}
\usepackage{subcaption}
\usepackage{multirow}

\usepackage{amsmath}
\usepackage{amsthm}
\usepackage{hyperref}
\usepackage{tabularx}

\usepackage[edges]{forest}
\forestset{
  dir tree/.style={
    for tree={
      parent anchor=south west,
      child anchor=west,
      anchor=mid west,
      inner ysep=-1pt,
      grow'=0,
      align=left,
      edge path={
        \noexpand\path [draw, \forestoption{edge}] (!u.parent anchor) ++(0.5em,-0.5em) |- (.child anchor)\forestoption{edge label};
      },
      font=\footnotesize,
      if n children=0{}{
        delay={
          prepend={[,phantom, calign with current]}
        }
      },
      fit=band,
      before computing xy={
        l=2em
      }
    },
  }
}

\renewcommand{\footnoterule}{%
	\kern -3pt
	\hrule
	\kern 2pt
}

\usepackage{microtype}
\usepackage{rotating}

\SetCommentSty{mycommfont}
\usepackage[sorting=none, mincitenames=1,backend=bibtex,doi=false,isbn=false,url=false]{biblatex} %
\usepackage{scalerel}
\usepackage{tikz} 

\usetikzlibrary{svg.path}

\definecolor{orcidlogocol}{HTML}{A6CE39}
\tikzset{
  orcidlogo/.pic={
    \fill[orcidlogocol] svg{M256,128c0,70.7-57.3,128-128,128C57.3,256,0,198.7,0,128C0,57.3,57.3,0,128,0C198.7,0,256,57.3,256,128z};
    \fill[white] svg{M86.3,186.2H70.9V79.1h15.4v48.4V186.2z}
                 svg{M108.9,79.1h41.6c39.6,0,57,28.3,57,53.6c0,27.5-21.5,53.6-56.8,53.6h-41.8V79.1z M124.3,172.4h24.5c34.9,0,42.9-26.5,42.9-39.7c0-21.5-13.7-39.7-43.7-39.7h-23.7V172.4z}
                 svg{M88.7,56.8c0,5.5-4.5,10.1-10.1,10.1c-5.6,0-10.1-4.6-10.1-10.1c0-5.6,4.5-10.1,10.1-10.1C84.2,46.7,88.7,51.3,88.7,56.8z};
  }
}

\newcommand\orcidicon[1]{\href{https://orcid.org/#1}{\mbox{\scalerel*{
\begin{tikzpicture}[yscale=-1,transform shape]
\pic{orcidlogo};
\end{tikzpicture}
}{|}}}}

\begin{document}

\onecolumn
\thispagestyle{empty}
\twocolumn
\setcounter{page}{1}
\setcounter{figure}{0}

\title{Bitwise Systolic Array Architecture for Runtime-Reconfigurable Multi-precision Quantized Multiplication on Hardware Accelerators
\vspace{-0.5em}
\vspace{5pt}
}

\author{
\IEEEauthorblockN{Yuhao Liu$^{1,3}$ \orcidicon{0000-0002-7281-2126}, \textit{Student Member, IEEE}, Salim Ullah$^{2}$ \orcidicon{0000-0002-9774-9522}, Akash Kumar$^{2,3}$ \orcidicon{0000-0001-7125-1737}, \textit{Senior Member, IEEE}}
\vspace{5pt}
\IEEEauthorblockA{$^{1}$Dresden University of Technology, Germany $^{2}$Ruhr University Bochum, Germany\\
$^{3}$Center for Scalable Data Analytics and Artificial Intelligence (ScaDS.AI Dresden/Leipzig), Germany\\
Email: yuhao.liu1@tu-dresden.de, \{salim.ullah, akash.kumar\}@rub.de}
}

\maketitle

\begin{abstract}
		 \input{arXiv/abstract}
\end{abstract}%

\thispagestyle{empty}


\section{Introduction}
\label{introduction}
\input{arXiv/introduction}

\section{Background}
\label{background}
\input{arXiv/background}

\section{Implementation}
\label{imp}
\input{arXiv/implementation}

\section{Evaluation}
\label{evaluation}
\input{arXiv/evaluation}

\section{Conclusion}
\label{concl}
\input{arXiv/conclusion}

\section{Acknowledgements}
\label{acknow}
\input{arXiv/acknowledgements}
\renewcommand{\bibfont}{\footnotesize}
\printbibliography

\end{document}

%% file: arXiv/abstract.tex
Neural network accelerators have been widely applied to edge devices for complex tasks like object tracking, image recognition, etc. Previous works have explored the quantization technologies in related lightweight accelerator designs to reduce hardware resource consumption. However, low precision leads to high accuracy loss in inference. Therefore, mixed-precision quantization becomes an alternative solution by applying different precision in different layers to trade off resource consumption and accuracy. Because regular designs for multiplication on hardware cannot support the precision reconfiguration for a multi-precision Quantized Neural Network (QNN) model in runtime, we propose a runtime reconfigurable multi-precision multi-channel bitwise systolic array design for QNN accelerators. We have implemented and evaluated our work on the \emph{Ultra96} FPGA platform. Results show that our work can achieve $1.3185\times$ to $3.5671\times$ speedup in inferring mixed-precision models and has less critical path delay, supporting higher clock frequency ($250MHz$).


%% file: arXiv/introduction.tex
Recent research of edge hardware devices widely applied~\emph{Neural Networks} (NN) on state-of-the-art applications, such as autonomous driving, the Internet of Things, wearable devices, voice and image recognition, etc. Considering the conflict between limited resources on the edge device and continually extending sizes of neural network models, related works explored the~\emph{Quantized Neural Network} (QNN) to reduce storage and hardware resource consumption by applying lower precision. For instance,~\emph{NVDLA}~\cite{Zhou2018} and~\emph{Vitis DPU}~\cite{vitis-ai} support the~\emph{INT8} 8-bit quantization in their deep learning processor designs.~\emph{FINN}~\cite{Umuroglu2017, Blott2018}, \emph{HLS4ML}~\cite{fahim2021hlsml}, \emph{LogicNets}~\cite{Umuroglu2020}, etc. proposed different frameworks to generate specialized inference accelerator designs on FPGA for the given low-precision trained ($<8$ bits) QNN models to reduce the on-chip resource consumption. However, various prior works have presented a higher accuracy loss in lower-precision quantized network models. For instance, the 1-bit quantized~\emph{Multilayer Perceptron} (MLP) model shown in work~\cite{Su2018} of Su et al. has an $8\times$ higher memory saving rate than the 8-bit quantized model applying the same network structure. However, the error of 1-bit models is about $32.7\%$ higher than the 8-bit model. Therefore, to trade off the low resource consumption and high accuracy loss in QNN hardware accelerator designs, the works from~\emph{HAQ}~\cite{mp3}, Chen et al.~\cite{mp1}, Tang et al.~\cite{mp2}, etc. explored mixed-precision quantization by using different precision in different layers. Compared to uniform quantization schemes across all layers (either high or low precision), mixed-precision quantized networks have middle-level inference accuracy and memory consumption for weight storage. We trained six tiny MLPs and six tiny~\emph{Convolution Neural Networks} (CNNs) models based on the~\emph{Brevitas}~\cite{brevitas} with different precision to evaluate the accuracy loss and memory saving. \emph{Tiny MLP} (TFC) and~\emph{Tiny CNN} (TCV) models are trained with the~\emph{MNIST} dataset~\cite{LeCun1998, deng2012mnist}. The TFC models comprise four layers with 64, 64, 64, and 10 neurons, respectively. TCV models have two convolution layers, each followed by a $2\times2$ max pooling layer. Following the final pooling layer are two fully connected layers. Each convolution layers have 64 $3\times3$ kernels, while two fully connected layers have 64 and 10 neurons, respectively. To achieve maximum compression of network weights, we apply lower precision to layers with a higher number of weights. Therefore, as shown in~\autoref{qnn_tab}, TFC applies 1/2/4/8-bit quantization, and TCV applies 4/1/2/8-bit quantization, respectively, as their mixed-precision schemes. Results show that 8-bit quantized models have the highest accuracy, similar to 32-bit floating-point-based networks. The two 1-bit models have the lowest accuracy with the least memory storage for weights. Meanwhile, two mixed-precision have balanced accuracy and memory requirements. 

\begin{table}[t]
    \centering
    \caption{Inference Accuracy of Quantized Network Models Applying Unified-Precision and Mixed-Precision Schemes}
    \resizebox{0.975\columnwidth}{!}
    {
        \begin{tabular}{clccccccc}
            \toprule
            \multicolumn{2}{c}{\multirow{2}{*}{Network Type}} & \multicolumn{7}{c}{Precision Settings in Four Layers of TFC and TCV models} \\
            \multicolumn{2}{c}{}                              & 1/1/1/1   & 2/2/2/2   & 1/2/4/8  & 4/1/2/8  & 4/4/4/4  & 8/8/8/8  & Float   \\ \midrule
            \multirow{2}{*}{TFC}  & Accuracy/\%    & 92.29     & 96.37     & 95.91    & -        & 97.55    & 97.36    & 97.89   \\
                                             & Weights/\textit{Byte}  & 7376      & 14752     & 9984     & -        & 29504    & 59008    & 236032  \\ \midrule
            \multirow{2}{*}{TCV}  & Accuracy/\%    & 96.26     & 98.96     & -        & 98.79    & 99.10    & 99.14    &  99.14  \\
                                             & Weights/\textit{Byte}  & 29848     & 59696     & -        & 55712    & 119392   & 238784   & 955136  \\ \bottomrule
        \end{tabular}
    }
    \label{qnn_tab}
\end{table}

\begin{table}[t]
    \centering
    \caption{Comparison of Unified- and Mixed-Precision Quantized MLPs Inferred on FPGA-based NN Accelerator}
    \resizebox{0.975\columnwidth}{!}
    {
        \begin{tabular}{cccccccl}
            \toprule
            \multirow{2}{*}{Design}    & \multirow{2}{*}{Precision} & \multirow{2}{*}{LUT}   & \multirow{2}{*}{FF}    & \multirow{2}{*}{BRAM} & \multirow{2}{*}{Frequency} & \multirow{2}{*}{Latency} & \multicolumn{1}{c}{\multirow{2}{*}{Accuracy}} \\ 
                                       &                            &                        &                        &                       &                            &                          & \multicolumn{1}{c}{}                          \\ \midrule
            \multirow{2}{*}{Vivado IP} & 8/8/8/8                    & \multirow{2}{*}{24090} & \multirow{2}{*}{22175} & \multirow{2}{*}{135}  & \multirow{2}{*}{150MHz}    & 137.654us                & 97.74\%                                       \\
                                       & 1/2/4/8                    &                        &                        &                       &                            & 131.059us                & 95.96\%                                      \\ \bottomrule
        \end{tabular}
    }
    \label{motivation_tab}
\end{table}

\begin{figure*}[t]
    \centering
    \centerline{\includegraphics[width=2\columnwidth]{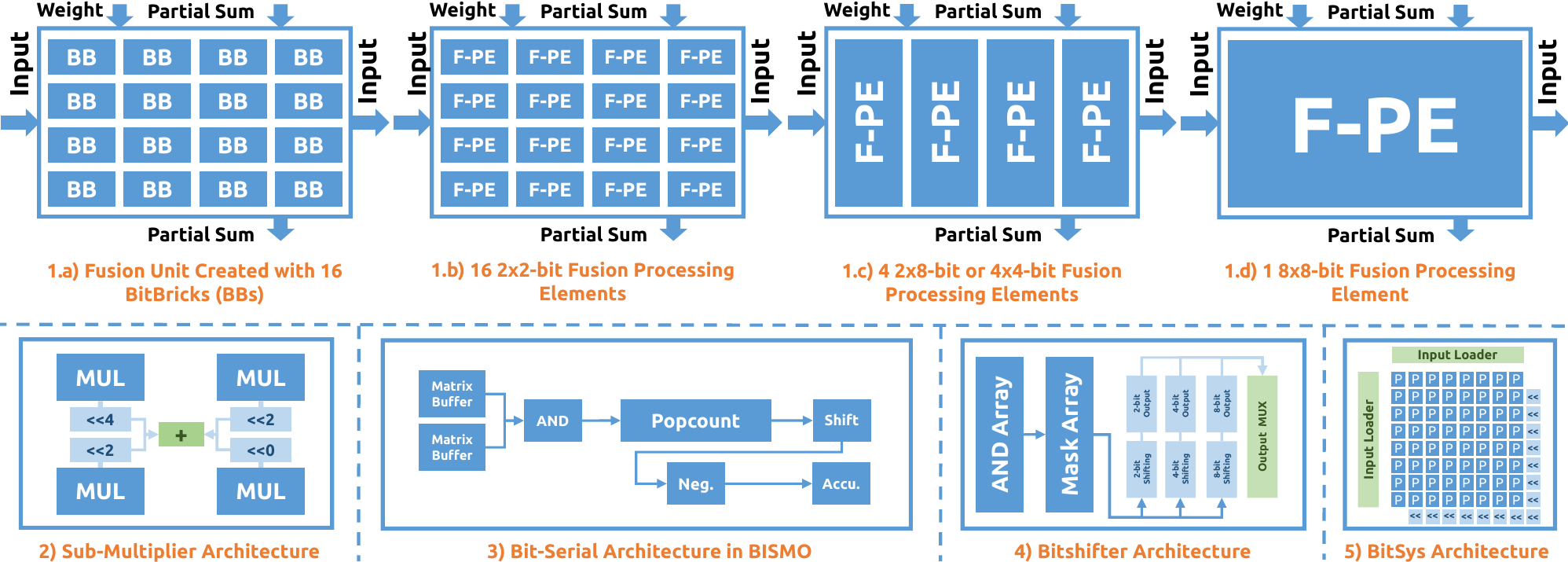}}
    \caption{Architectures of Prior Works and \emph{BitSys}}
    \label{related_works_fig}
\end{figure*} 

\subsection{Motivation}

Mixed-precision QNNs show the potential to achieve a better and more flexible trade-off between resource consumption and accuracy loss. Prior works explored the design of related accelerators better to support the inference of mixed-precision networks on hardware. Results report that utilizing fixed-precision multipliers diminishes the performance advantages of mixed-precision accelerators. As shown in~\autoref{motivation_tab}, one previous work of Liu et al.~\cite{Liu2023} implemented one single-layer NN accelerator on~\emph{Ultra96-V2} FPGA platform with 64 8-bit integer~\emph{Vivado} multiplier IPs to infer one 8-bit quantized MLP and one 1/2/4/8-bit mixed-precision quantized MLP trained by~\emph{Brevitas}~\cite{brevitas} with~\emph{MNIST} dataset~\cite{LeCun1998, deng2012mnist}. Both MLPs have four layers with 64, 64, 64, and 10 neurons, respectively. \autoref{motivation_tab} listed the average inference latency of one~\emph{MNIST} input, computed by averaging the total latency of 1000 times inputs. The results indicate that the
inference speed of the mixed-precision MLP has not significantly improved compared to the uniformly 8-bit quantized network. Because the input width of 8-bit~\emph{Vivado} multiplier IP cannot be reconfigured as 1/2/4-bit in runtime, all input data must be unified and extended to the largest precision, 8 bits. As a result, the inference acceleration of the mixed-precision model can only benefit from the transmission speedup between off-chip memory and FPGA based on low-precision data, not from the computation. Therefore, if multipliers can reconfigure the input precision and channel number in runtime, for instance, reset a single-channel 8-bit input as a dual-channel 4-bit input for signed 8/4-bit quantized layers, the inference of mixed-precision network models can be sped up on hardware.

\subsection{Contributions}

Prior works, such as \emph{PIR-DSP}~\cite{PIR-DSP}, \emph{BitFusion}~\cite{bitfusion}, \emph{Multiplier-Tree}~\cite{Liu2023}, \emph{Bitshifter}~\cite{Liu2023}, etc., explored the designs of multi-precision multipliers. Extending on our abstract in~\cite{bitsys}, we proposed a~\emph{Bitwise Systolic Array Architecture} (BitSys) in this manuscript supporting quantized multi-precision multi-channel runtime reconfigurable multiplication for neural network accelerator designs. The key features and contributions of this work are:\footnote{The RTL source code, Vivado projects, and packaged Vivado IPs for the BitSys multiplier and MAC are available at \url{https://github.com/liuyh-Horizon/BitSys} to facilitate validation, reproduction, and reuse.}

\begin{itemize}
    \item We implemented one systolic-array-based multiplier, \emph{BitSys}, based on the bitwise (1-bit) processing element and optimized it with LUT primitive for FPGA. Our design supports runtime reconfiguration for signed/unsigned 8/4/2/1-channel 1/2/4/8-bit multiplication. Moreover, this multiplier is specially designed to support the XNOR multiplication for the~\emph{Binarized Neural Network} (BNN) in~\emph{FINN}~\cite{Umuroglu2017, Blott2018}.

    \item We extended our multiplier as a Multiply-Accumulator (MAC) to implement one single-layer accelerator and one systolic array accelerator and evaluate them for the mixed-precision model inference acceleration.
\end{itemize}

We evaluated our multipliers, MAC, and accelerator implementations on the \emph{Ultra96-V2} FPGA platform and compared them with previous works. The synthesis and implementation report in~\emph{Vivado} shows our designs have low critical path delay from $1.357ns$ to $1.719ns$. The measurement result proves that our systolic array accelerator is $1.3185\times$ to $3.5671\times$ faster in the inference of mixed-precision networks than previous works.

\subsection{Organization}

This manuscript is structured as follows: Section II compares our \emph{BitSys} design with related works. Section III introduces implementations of \emph{BitSys} architecture. Section IV shows the evaluation results on~\emph{Ultra96-V2} platform compared with related works. Section V concludes the contents of this paper.

%% file: arXiv/background.tex
\subsection{Classification of Prior Multi-precision Multiplier Designs}

\begin{table*}[t]
    \centering
    \caption{Differences between the BitSys Architecture and Previous Works}
    \resizebox{0.975\linewidth}{!}
    {
        \begin{tabular}{cccccccccccccccc}
        \toprule
        \multirow{2}{*}{Work} & \multirow{2}{*}{Platform} & Accu. or & No & Signed or & \multicolumn{10}{c}{Available  Precision}                                                     &         \\
                              &                           & Approx.  & DSP                     & Unsigned  & $1 \times 1$ & $2 \times 2$ & $4 \times 4$ & $4 \times 16$ & $8 \times 8$ & $8 \times 16$ & $9 \times 9$ & $16 \times 16$ & $18 \times 27$ & $24 \times 24$ & $32 \times 32$ \\ \midrule
        Guo et al.~\cite{guo2020}            & FPGA                      & Approx.  & $\surd$                    & Signed    & $\times$     & $\times$     & $\times$     & $\surd$      & $\times$     & $\surd$      & $\times$     & $\times$       & $\times$       & $\times$       & $\times$       \\
        Neda et al.~\cite{neda2022multi}           & FPGA                      & Approx.  & $\surd$                    & Signed    & $\times$     & $\times$     & $\times$     & $\times$      & $\surd$     & $\times$      & $\times$     & $\surd$       & $\times$       & $\times$       & $\times$       \\
        Shun et al.~\cite{mm1}           & FPGA                      & Accu.    & $\surd$                    & Signed    & $\times$     & $\times$     & $\times$     & $\times$      & $\surd$     & $\times$      & $\times$     & $\surd$       & $\times$       & $\surd$       & $\surd$       \\
        Pfänder et al.~\cite{mm2}        & FPGA                      & Accu.    & $\surd$                    & Both      & $\times$     & $\times$     & $\times$     & $\times$      & $\surd$     & $\surd$      & $\times$     & $\surd$       & $\times$       & $\surd$       & $\surd$       \\
        PIR-DSP~\cite{PIR-DSP}               & FPGA                      & Accu.    & $\times$                    & Both      & $\times$     & $\surd$     & $\surd$     & $\times$      & $\times$     & $\times$      & $\surd$     & $\times$       & $\surd$       & $\times$       & $\times$       \\
        Multiplier-Tree~\cite{Liu2023}       & FPGA                      & Accu.    & $\surd$                    & Both      & $\surd$     & $\surd$     & $\surd$     & $\times$      & $\surd$     & $\times$      & $\times$     & $\surd$       & $\times$       & $\times$       & $\surd$       \\
        Bitshifter~\cite{Liu2023}            & FPGA                      & Accu.    & $\surd$                    & Both      & $\surd$     & $\surd$     & $\surd$     & $\times$      & $\surd$     & $\times$      & $\times$     & $\surd$       & $\times$       & $\times$       & $\surd$       \\ \midrule
        \textbf{BitSys (Ours)}         & \textbf{FPGA}                      & \textbf{Accu.}    & \textbf{$\surd$}                    & \textbf{Both}      & \textbf{$\surd$}     & \textbf{$\surd$}     & \textbf{$\surd$}     & \textbf{$\times$}      & \textbf{$\surd$}     & \textbf{$\times$}      & \textbf{$\times$}     & \textbf{$\times$}       & \textbf{$\times$}       & \textbf{$\times$}       & \textbf{$\times$}       \\ \bottomrule
        \end{tabular}
    }
    \label{diff_bitsys_tab}
\end{table*}

Previous work explored different schemes for multi-precision multiplier designs, which can be classified by bit-serial/bit-parallel architectures and fixed/variable input widths. 

Bit-serial multipliers execute the bitwise processing for multiplication in serial, such as~\emph{BISMO}~\cite{Umuroglu2018}, the work of Ienne et al.~\cite{bitserial2}, the work of Shafer et al.~\cite{bitserial1}, etc. For example, as shown in~\autoref{related_works_fig}.3, \emph{BISMO} loads the inputs with the batch size of $k$-bit to execute the pipelined processing in serial. For $m$-bit inputs, it takes $\frac{m}{k}$ clock cycles to complete the multiplication. As a result, low-precision multiplication consumes fewer clock cycles than high-precision. Therefore, this design scheme can support temporal reconfiguration for different precision in runtime by completing more multiplications for lower precision in $m$ clock cycles. However, for $n$ times inputs, this scheme requires $n \times m$ cycles in computation, which leads to a high inference latency in hardware accelerators. 

Therefore, most prior works are designed as bit-parallel architectures based on sub-multiplier schemes as shown in~\autoref{related_works_fig}.2, which generate one output per clock cycle, such as the works of Neda et al.~\cite{neda2022multi}, Guo et al.~\cite{guo2020}, Liu et al.~\cite{Liu2023}, Pfänder et al.~\cite{mm2}, and~\emph{PIR-DSP}~\cite{PIR-DSP}. For $2n\times2n$-bit multiplication, $A \times B = A_0 B_0 \times 2^{2n} + ( A_1 B_0 + A_0 B_1 ) \times 2^{n} + A_1 B_1$, if two inputs are split as four $n$-bit data, $A_0$, $A_1$, $B_0$, and $B_1$, the multiplication result is computed by summing the products of four $n \times n$-bit sub-multiplier results by $2^{2n}$, $2^{n}$, and $1$ separately, which can be converted as $2n$/$n$/0-bit preset left-shifting. Therefore, if we bypass the outputs of two sub-multipliers with $n$-bit left shifting, the sum of four sub-multipliers is dual-channel $n \times n$-bit multiplication. Otherwise, the result is single-channel $2n \times 2n$-bit multiplication. 

However, bypassing two sub-multipliers leads to low hardware efficiency. The works of Li et al.~\cite{BSC1}, Dai et al.~\cite{BSC2}, and~\emph{BitFusion}~\cite{bitfusion} explored another scheme to utilize all sub-multipliers in different precision. For instance, the~\emph{BitFusion} architecture shown in~\autoref{related_works_fig}.1 implemented sixteen 2-bit multipliers, \emph{BitBricks} (BBs), as the basic processing elements, \emph{F-PE}, to organize a systolic array. Based on the principle of sub-multiplier architecture designs, sixteen 2-bit multipliers in \emph{BitFusion} can create four 4-bit multipliers and one 8-bit multiplier. The major difference is, as shown in~\autoref{related_works_fig}.1c, \emph{BitFusion} applies the reconfigurable, not preset, left-shifters. Therefore, four 2-bit multipliers created a large \emph{F-PE} to support both $2\times8$-bit and $4\times4$-bit multiplications to utilize all BBs with different input widths as 10 and 8 bits. The variable input width complicates the data streaming control designed as a series of multiplexers and registers. In principle, the~\emph{BitFusion} presents a multi-precision systolic array, not a multi-precision multiplier. Only the F-PE in in~\autoref{related_works_fig}.1c is the reconfigurable multiplier. For instance, as shown in~\autoref{related_works_fig}.1b, c, and d, \emph{BitFusion} works as a $4\times4$, $1\times4$, $1\times1$ systolic array separately. This design limited the scenario of~\emph{BitFusion} architecture as the tensor processing unit.

Differing from the designs mentioned above, Liu et al.~\cite{Liu2023} proposed a~\emph{Bitshifter} architecture inspired by the~\emph{BISMO}~\cite{Umuroglu2018} converting the multiplication as the combination of bitwise AND and left-shifting. This is a bit-parallel multi-precision multiplier with a fixed input width. The result of $N$-bit multiplication, $A \times B=\sum_{i=0}^{n-1}\sum_{j=0}^{n-1} 2^{i+j}a_{i}b_{j}$, is the sum of $2^{i+j}a_{i}b_{j}$. $a_{i}$ and $b_{j}$ are the bit values of $A$ and $B$, $a_{i}b_{j}$ is the bitwise AND, and $2^{i+j}$ can be converted as the preset left-shifting. Therefore, as shown in~\autoref{related_works_fig}.5,~\emph{Bitshifter} architecture computes all $a_{i}b_{j}$ with bitwise AND first, then filters the unnecessary results with the mask for different precision and applies the corresponding left-shifting to compute partial products.

\subsection{Comparison between the BitSys and Previous Works}

Considering the motivation in~\autoref{introduction}A, we target to explore a multi-precision multiplier design to speed up the computation of mixed-precision QNN models on hardware. To this end, we exclude the bit-serial multiplier scheme in our scope because of its long computation latency. To simplify the data steaming control and deploy our multiplier in variable scenarios of the existing hardware designs, such as the systolic array, single-layer accelerator, etc., we have not adopted the architecture similar to the~\emph{BitFusion} and works of Li et al.~\cite{BSC1} and Dai et al.~\cite{BSC2}. Therefore, our \emph{BitSys} architecture presented a bit-parallel and input-width-fixed multi-precision multiplier design, inspired by~\emph{BitShifter}~\cite{Liu2023} and~\emph{BitFusion}~\cite{bitfusion} by converting the multiplication with bitwise operation with partial product mask and computing them with a systolic array.~\autoref{diff_bitsys_tab} compared it with related works. In this table, $\surd$ and $\times$ mean the selected features, like available precision, are applied in the corresponding works or not:  

\begin{itemize}
    \item Both two inputs of \emph{BitSys} support multi-channel reconfiguration for variable precision. The work of Guo et al.~\cite{guo2020} only supports 1/2-channel $2N/N \times M$-bit multiplication. 
    \item Our work supports accurate computing, not the approximate designs of Neda et al.~\cite{neda2022multi} and Guo et al.~\cite{guo2020}. 
    \item Shun et al.~\cite{mm1} proposed an accurate multi-precision multiplier based on~\emph{Radix-4 Booth} multiplier. However, it is designed for 8/16/24/32-bit multiplication, which is unsuitable for the 1/2/4/8-bit multiplication we targeted for low-precision QNN models. 
    \item Pfänder et al.~\cite{mm2} extended the work of Shun et al.~\cite{mm1} as serial processing to reduce resource consumption. In contrast to this work, \emph{BitSys} adopts the bit-parallel architecture to speed up computation in hardware accelerators.
    \item \emph{PIR-DSP}~\cite{PIR-DSP} focuses on designing multi-precision multipliers based on DSP slices of FPGA. However, the input widths of \emph{DSP48}/\emph{DSP58} resources in \emph{Xilinx} FPGA are wider than 1/2/4/8-bit QNN models. Meanwhile, DSP slices cannot process the XNOR multiplication in BNN. Therefore, DSP slices are inefficient and unsuitable in designing our \emph{BitSys} architecture. 
    \item As shown in~\autoref{related_works_fig}.4 and~\autoref{related_works_fig}.5, we fused the AND array and Mask array in \emph{Bitshifter}~\cite{Liu2023} as a bitwise systolic array inspired by \emph{BitFusion}~\cite{bitfusion} for higher throughputs. The left shifters and output generation stages for different precision are fused as output generation pipelines in \emph{BitSys}.  
    \item The processing elements in \emph{BitSys} execute 1-bit operations, supporting higher clock frequency with lower critical path delay. Moreover, the XNOR multiplication in \emph{Bitshifter}~\cite{Liu2023} is computed in an individual module. We fused it in our 1-bit processing elements to save the hardware resources. 
    \item We implemented the single-layer accelerator and systolic array accelerator based on \emph{BitSys} to show its potential to be applied in different designs.  
\end{itemize}

\begin{figure}[t]
    \centering
    \begin{minipage}{\columnwidth}
        \centerline{\includegraphics[width=\columnwidth]{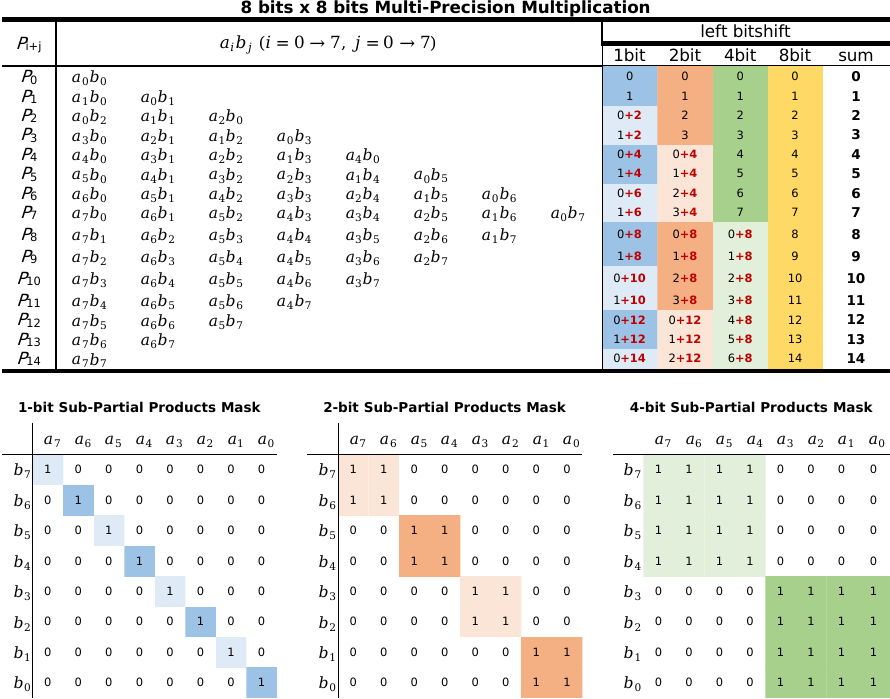}}
        \caption{1/2/4/8 Channels 8/4/2/1 bits Multiplication and Corresponding Partial Products Masks}
        \label{bit_table}
    \end{minipage}
    \begin{minipage}{\columnwidth}
        \centerline{\includegraphics[width=\columnwidth]{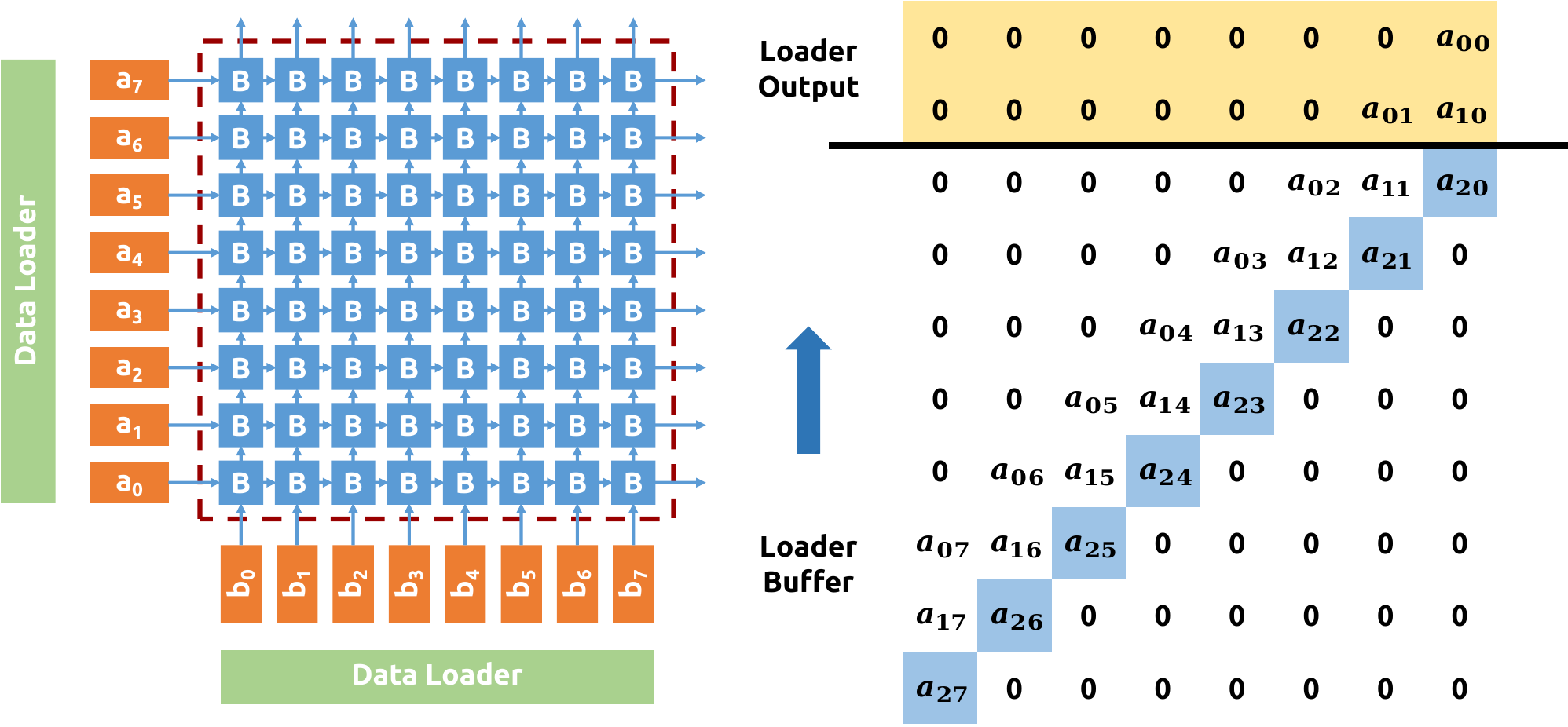}}
        \caption{Bitwise Systolic Array (left) and Input Loader (right)}
        \label{bit_loader}
    \end{minipage}
\end{figure}

%% file: arXiv/implementation.tex
\subsection{Mathematics Principle}

For $N$-bit multiplication, $A \times B=\sum_{i=0}^{n-1}\sum_{j=0}^{n-1} 2^{i+j}a_{i}b_{j}$, ($0\leqslant i$, $ j \leqslant n-1$), $a_{i}$ and $b_{j}$ are the bit value of $A$ and $B$, $2^{i+j}$ can be replaced as left shifting, $\ll(i+j)$, and $a_{i}b_{j}$ is the bitwise $AND$. As shown in~\autoref{bit_formular}, if we define $a_{i}b_{j}$ as sub-partial products, for $N$-bit multiplication, we can reorganize all sub-partial products as $2N-1$ groups. Each group applies the same left-shifting value $M$, ($0\leqslant M \leqslant 2N-2$). Therefore, we can define the sum of one group as the partial products, $P_{M}=\sum_{M=i+j}a_{i}b_{j}$. Therefore, all $a_{i}b_{j}$ are organized as the triangular-aligned structure shown in~\autoref{bit_table}. The bold parts in~\autoref{bit_formular} represent the sign bit with $\pm$ in the multiplication. When it is negative, the multiplication is signed. Therefore, by switching to add or subtract the AND results of $a_{k}b_{n-1}$ and $a_{n-1}b_{k}$ ($0\leqslant k < n-1$) from partial products, the multiplication can be reconfigured as signed/unsigned computing. After applying the corresponding left-shifting value for each partial product, their sum is the product of $N$-bit multiplication.

Based on the basic mathematics principle mentioned above, as shown in~\autoref{bit_table}, we extend it for runtime reconfigurable multi-channel multi-precision multiplication: Using $8 \times 8$-bit multi-precision multiplication as an example, $P_{i+j}$ ($0\leqslant i \leqslant 7$, $0\leqslant j \leqslant 7$) are the partial products in this computation, which are the sum of corresponding sub-partial products, $a_{i}b_{j}$, shown in the same row of $P_{i+j}$ in the second column of~\autoref{bit_table}. For different precision, three sub-partial product masks shown in~\autoref{bit_table} select the desired sub-partial products, $a_{i}b_{j}$, for 8/4/2 channel 1/2/4-bit multiplications. For instance, for the dual-channel $4 \times 4$-bit multiplication, two green squares in the 4-bit sub-partial products mask of~\autoref{bit_table} select the desired $a_{i}b_{j}$ in computation. The filtered $a_{i}b_{j}$ are set as zero, and one green square in the 4-bit sub-partial products mask selects the $a_{i}b_{j}$ for one channel. Based on the same principle, the four orange and eight blue squares in 2/1-bit masks select the desired sub-partial products for corresponding 4/8 channels. All $a_{i}b_{j}$ are used to compute single-channel $8 \times 8$-bit multiplication. Therefore, we can compute all sub-partial products first, reconfigure the mask in runtime to filter the undesired $a_{i}b_{j}$ for different precision and channels, and then compute the sum of filtered $a_{i}b_{j}$ as partial products, $P_{i+j}$. Considering the lower hardware utilization efficiency when more sub-partial products are filed as zero in lower precision, compared with the disabled sub-multipliers in the previous works shown in the~\autoref{related_works_fig}.2 of~\autoref{background}, this is a common trade-off to achieve the multi-precision reconfiguration for related bit-parallel input-width-fixed multiplier designs.

\begin{equation}
    \small
    \begin{aligned}
        A \times B = & \langle a_{n-1}a_{n-2}...a_{1}a_{0}\rangle_{bin} \times \langle b_{n-1}b_{n-2}...b_{1}b_{0}\rangle_{bin} \\
        = & (\mathbf{\pm2^{n-1}a_{n-1}}+2^{n-2}a_{n-2}+...+2^{1}a_{1}+2^{0}a_{0}) \\
          & \times (\mathbf{\pm2^{n-1}b_{n-1}}+2^{n-2}b_{n-2}+...+2^{1}b_{1}+2^{0}b_{0})\\
        = & [(a_{n-1}b_{n-1})\ll2n-2] \\
        + & [(\mathbf{\pm a_{n-1}b_{n-2}}+\mathbf{\pm a_{n-2}b_{n-1}})\ll2n-3]\\
        + & ... \\
        + & [(a_{1}b_{0}+a_{0}b_{1})\ll1]\\
        + & [(a_{0}b_{0})\ll0]
    \end{aligned}
    \label{bit_formular}
\end{equation}

After we get the value of partial products, $P_{i+j}$, the multiplier needs to apply the corresponding left shifting to $P_{i+j}$ and sum them as the multi-channel results. Therefore, as shown in the~\emph{left bitshift} column of~\autoref{bit_table}, for example, when the multiplier executes 8-channel 1-bit multiplication, each channel needs two partial products and applies 0/1-bit left shifting separately. For instance, the result in the first channel of 1-bit multiplication is $(P_{0}\ll 0)+(P_{1}\ll 1)$. Actually, the 1-bit operation only needs one $P_i$ in each channel, such as $P_0$ for the first channel. However, to keep the output as 8-channel-2-bit, the $P_1$ is used as a placeholder, and its $a_{1}b_{0}$ and $a_{0}b_{1}$ are filtered as 0 by 1-bit sub-partial product masks. Based on the same principle, for instance, we can infer that $P_3$ is also a placeholder partial product for the 1st channel of 2-bit multiplication. Considering the total output width of this 8-bit multiplier is 16 bits, the output widths of one channel in 1/2/4-bit modes are 2/4/8 bits. Therefore, in the final output, each 2/4/8-bit output from the $i$-th channel in 1/2/4-bit multiplication needs a channel offset by left-shifting to $(i-1)\times2$, $(i-1)\times4$, and $(i-1)\times8$ bits to avoid conflict with the $(i-1)$-th channel. As shown in the \emph{sum} of \emph{left bitshift} column in~\autoref{bit_table}, for each partial product, $P_{k}$ ($0\leqslant k \leqslant 14$), the sums of partial product left shifting (black numbers in~\emph{left bitshift} column) and channel offset left shifting (red numbers in~\emph{left bitshift} column) are always $k$ in all 1/2/4/8-bit multiplication modes. Therefore, differing from the individual three left shifting stages in \emph{Bitshifter} architecture~\cite{Liu2023} shown in~\autoref{related_works_fig}.4, our \emph{BitSys} applied the same left shifting for each partial product in all 1/2/4/8-bit modes. In conclusion, the computation of the runtime reconfigurable multi-precision multiplication in our work can be converted into four steps: 

\begin{enumerate}
    \item Computing all $a_{i}b_{j}$ ($0\leqslant i \leqslant n-1$, $0\leqslant j \leqslant n-1$).
    \item Filtering to get the desired $a_{i}b_{j}$ with corresponding sub-partial products mask for different precision.
    \item Computing the partial products, $P_{k}$ ($0\leqslant k \leqslant 2n-2$), and applying $k$-bit left shifting.
    \item Computing the sum of $P_{k}$ as the final output.
\end{enumerate}

\begin{figure}[t]
    \centering
    \begin{minipage}{\columnwidth}
        \centerline{\includegraphics[width=\columnwidth]{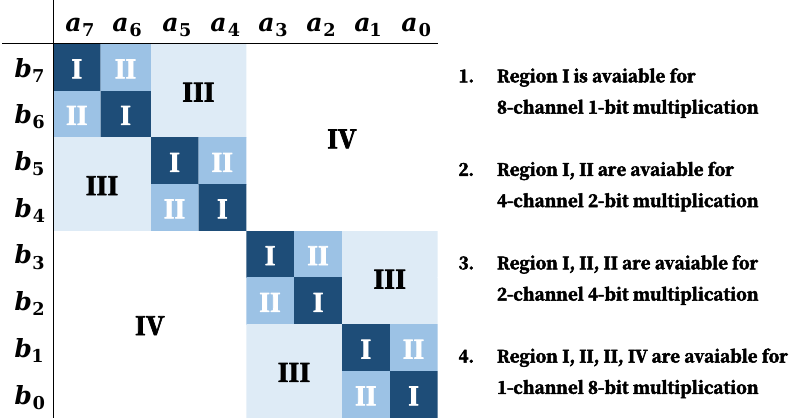}}
        \caption{Bitwise Processing Element Location in Systolic Array}
        \label{bit_regions}
        \vspace{7pt}
    \end{minipage}
    \begin{minipage}{\columnwidth}
        \centerline{\includegraphics[width=\columnwidth]{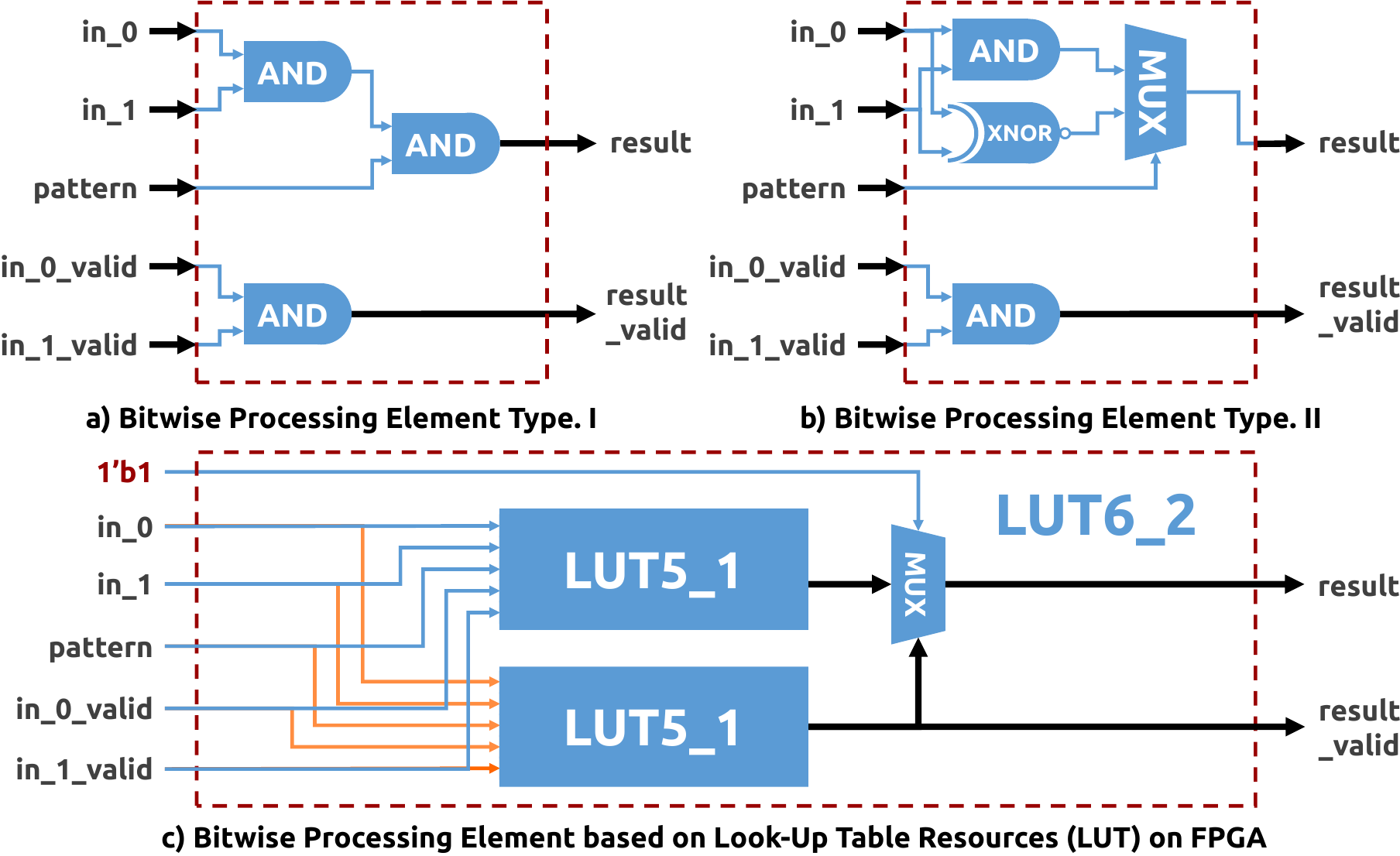}}
        \caption{Design of Bitwise Processing Element}
        \label{lut_structure}
    \end{minipage}
\end{figure} 

\subsection{Bitwise Systolic Array Architecture for Multi-precision Multiplier}
    
\begin{figure*}[t]
    \centering
    \begin{minipage}{0.42\textwidth}
        \centerline{\includegraphics[width=\columnwidth]{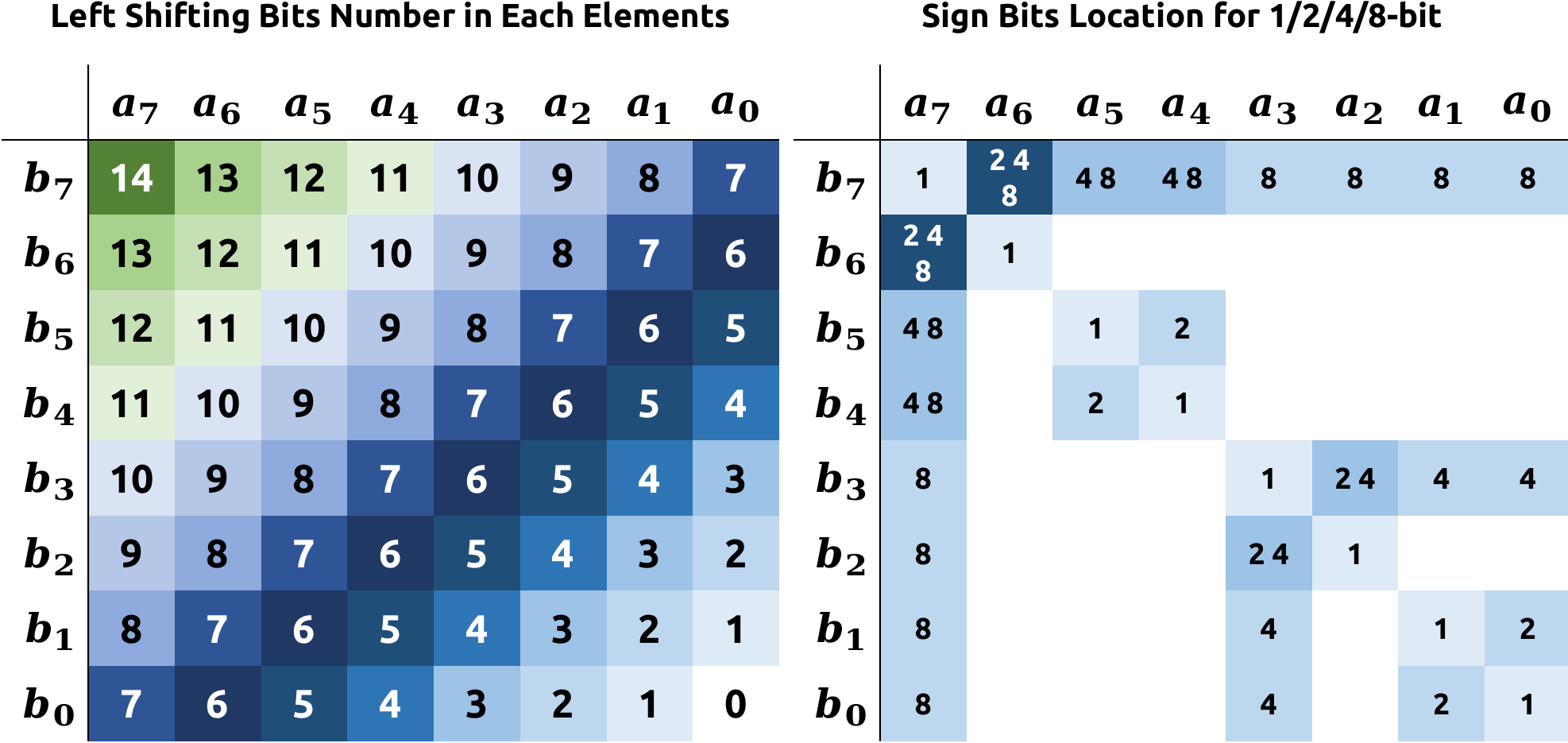}}
        \caption{Left-Shifting of Diagonal \& Signed Elements}
        \label{diagonal}
    \end{minipage}
    \begin{minipage}{0.55\textwidth}
        \centerline{\includegraphics[width=\columnwidth]{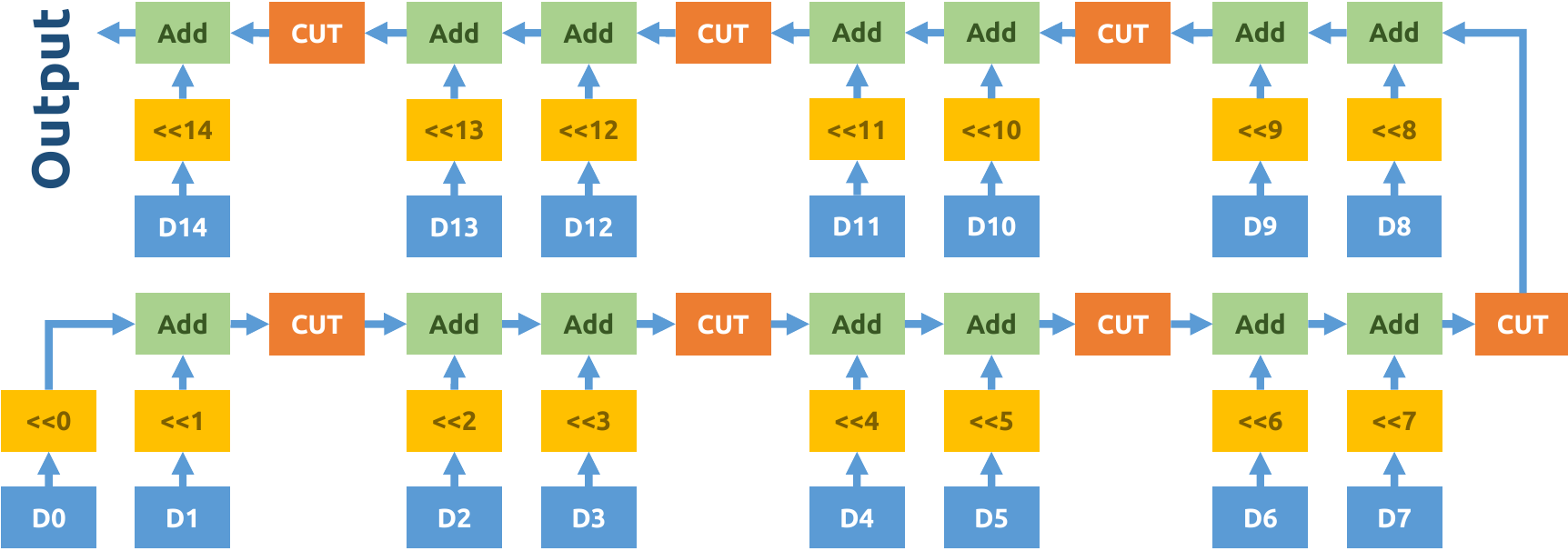}}
        \caption{Design of Output Generator Pipeline}
        \label{diagonal_pip}
    \end{minipage}
\end{figure*} 

\begin{figure}[t]
    \centering
    \centerline{\includegraphics[width=0.95\columnwidth]{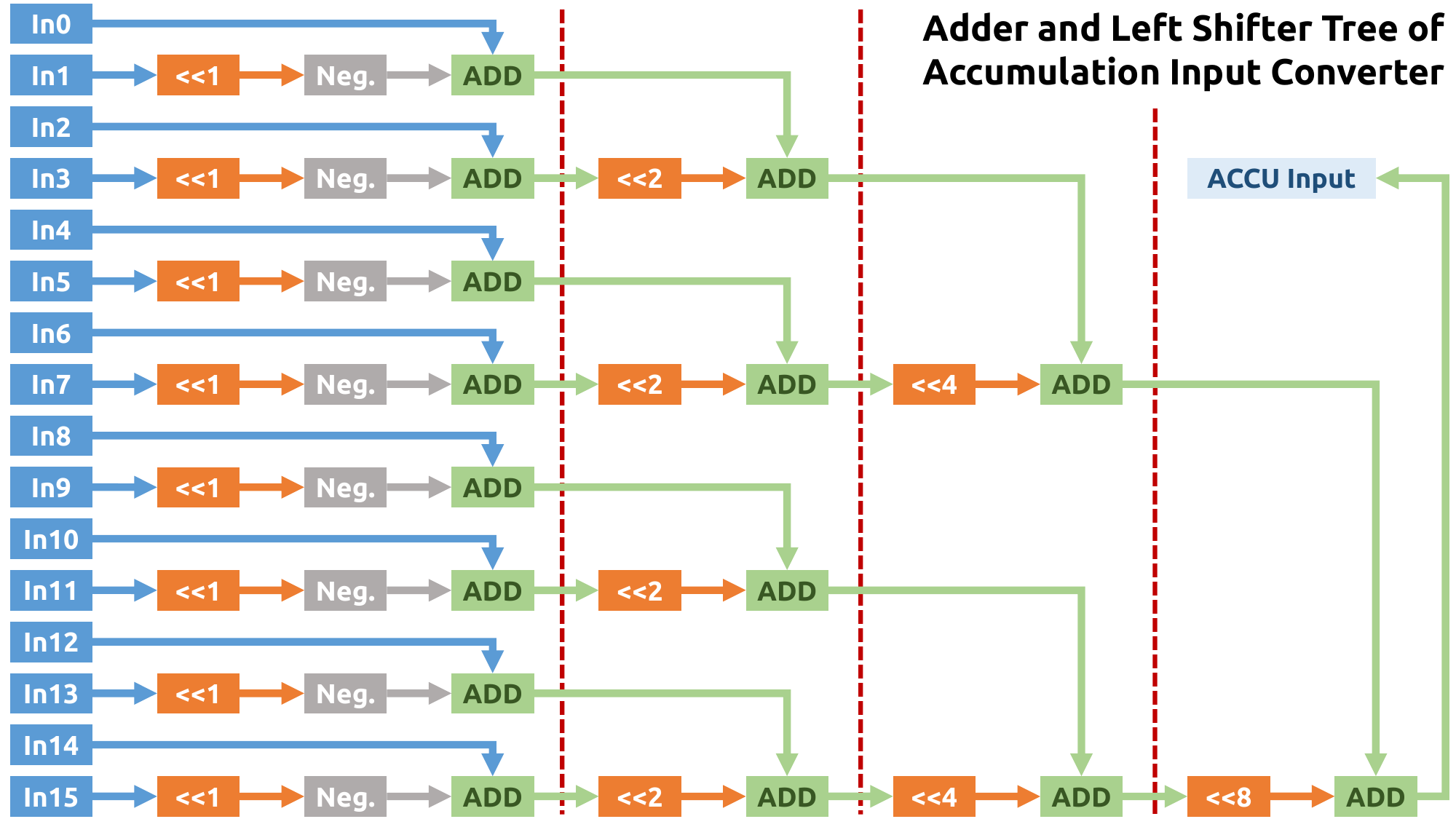}}
    \caption{Multi-Precision Accumulator Input Converter}
    \label{accu_structure}
\end{figure}

To execute the first two steps mentioned above, we implemented a bitwise systolic array as shown in~\autoref{bit_loader} (left) and the input loader as shown in~\autoref{bit_loader} (right). The input loader works to prepare the inputs for the bitwise systolic array. For instance, in an 8-bit multiplier, the input loader implements a tiny FIFO buffer with eight $8$-bit registers, loading one new input in the diagonal (blue bits) and pushing the data from the bottom to the top as the loader outputs (yellow part). One bitwise systolic array requires two input loaders. The bitwise systolic array we implemented in \autoref{bit_loader} (left) consists of bitwise processing elements, which fused the sub-partial products mask and 1-bit arithmetic operations. Considering the multiplication in the BNN presented in FINN~\cite{Umuroglu2017, Blott2018} is the XNOR operation, which represents the -1 as \emph{'0'} and +1 as \emph{'1'}, we need two kinds of bitwise processing elements as shown in~\autoref{lut_structure}a and~\autoref{lut_structure}b: Type.I element switches between 1-bit XNOR and AND operation for 1-bit or 2/4/8-bit multiplication. Type.II element switches between 1-bit AND operation and zero output according to the sub-partial product mask in variable precision. \autoref{bit_regions} presents the location mapping of bitwise processing elements and when they are available for different precision according to the sub-partial product masks: Type.I elements are located in \emph{Region I} and Type.II elements are located in \emph{Region II/III/IV}. For instance, when the multiplier works in 1-bit mode, the processing elements in \emph{Region I} output the results of 1-bit XNOR, and other elements output 0. When precision is 4-bit, the processing elements in \emph{Region I/II/III} output the results of 1-bit AND, and other elements output 0. One pattern signal generated according to the current precision controls the output switching of one bitwise processing element. Furthermore, because when two inputs of the 1-bit XNOR are '0', the output is '1', the bitwise processing element needs input and output a valid signal for the following adder to avoid mistake accumulation when no input is loaded. Therefore, as shown in~\autoref{lut_structure}c, we define one bitwise processing element as 6-bit input and 2-bit output module: 2-bit input, 2-bit input valid, 1-bit pattern switching, 1-bit input is always '1' to enable 2-bit output, 1-bit output, and 1-bit output valid signal. Therefore, one bitwise processing element for both types can be implemented as one \emph{LUT6\_2} primitive in \emph{Xilinx} FPGA. 


For the second two steps in the computation of our \emph{BitSys}, the multiplier needs to compute the value of partial products, $P_{k}$ ($0\leqslant k \leqslant 2n-2$), apply the left-shifting to them, and add all $P_{k}$ as the final output. As shown in~\autoref{diagonal} (left), numbers in this figure are the left-shifting bits applied to the outputs of their located bitwise processing elements. Therefore, the sum of the bitwise processing element results with the same left-shifting bits, which are in the same diagonal, is a partial product. Considering the signed multiplication in~\autoref{bit_formular}, the numbers in~\autoref{diagonal} (right) represent that, in which precision, the outputs of bitwise processing elements they located need to be subtracted from partial products. For instance, $a_{7}b_{6}$ needs to be subtracted in 2/4/8-bit multiplication because $a_{7}$ is a sign bit in this precision. $a_{7}b_{7}$ needs to be subtracted in 1-bit multiplication because the XNOR output is signed output, representing -1 as '0' and +1 as '1'. Moreover, because both $a_{7}$ and $b_{7}$ are sign bits in 2/4/8-bit multiplication, $a_{7}b_{7}$ does not need to be subtracted. After finishing the computation of partial products, $P_{k}$, our \emph{BitSys} multiplier loads them as the inputs, $D_{k}$, of the output generator pipeline shown in~\autoref{diagonal_pip} to apply $k$-bit left-shifting and sum the left-shifted partial products as final output. Considering the sum of signed partial products generates the carry bits in computation and influences the result in the next channel, we insert the carry-cutter modules in the output generator pipeline to limit the output width. For instance, in 1-bit multiplication, all carry-cutters are enabled to limit the output width of 8 channels; in 2-bit multiplication, only the carry-cutters after $D_{3,7,11}$ are enabled to limit the output width of 4 channels. Because the bitwise systolic array generates the partial product from $P_{0}$ to $P_{14}$ sequentially and executes multiple computations simultaneously, our output generator pipeline is designed for pipelined parallel processing. For instance, in the 1st cycle, the bitwise systolic array outputs the $D_0$ of $MUL_0$, and the output generator pipeline left-shifts it to 0-bit. In the 2nd cycle, the bitwise systolic array outputs the $D_0$ of $MUL_1$ and $D_1$ of $MUL_0$. The output generator pipeline applies the 0/1-bit left-shifting on them separately and adds the $D_0$ and 1-bit left-shifted $D_1$ of $MUL_0$ together for the next step.

\subsection{Single-Layer and Systolic Array Accelerator Implementation based on \emph{BitSys}}
\begin{figure*}[t]
    \centering
    \begin{minipage}{0.385\textwidth}
        \centerline{\includegraphics[width=\columnwidth]{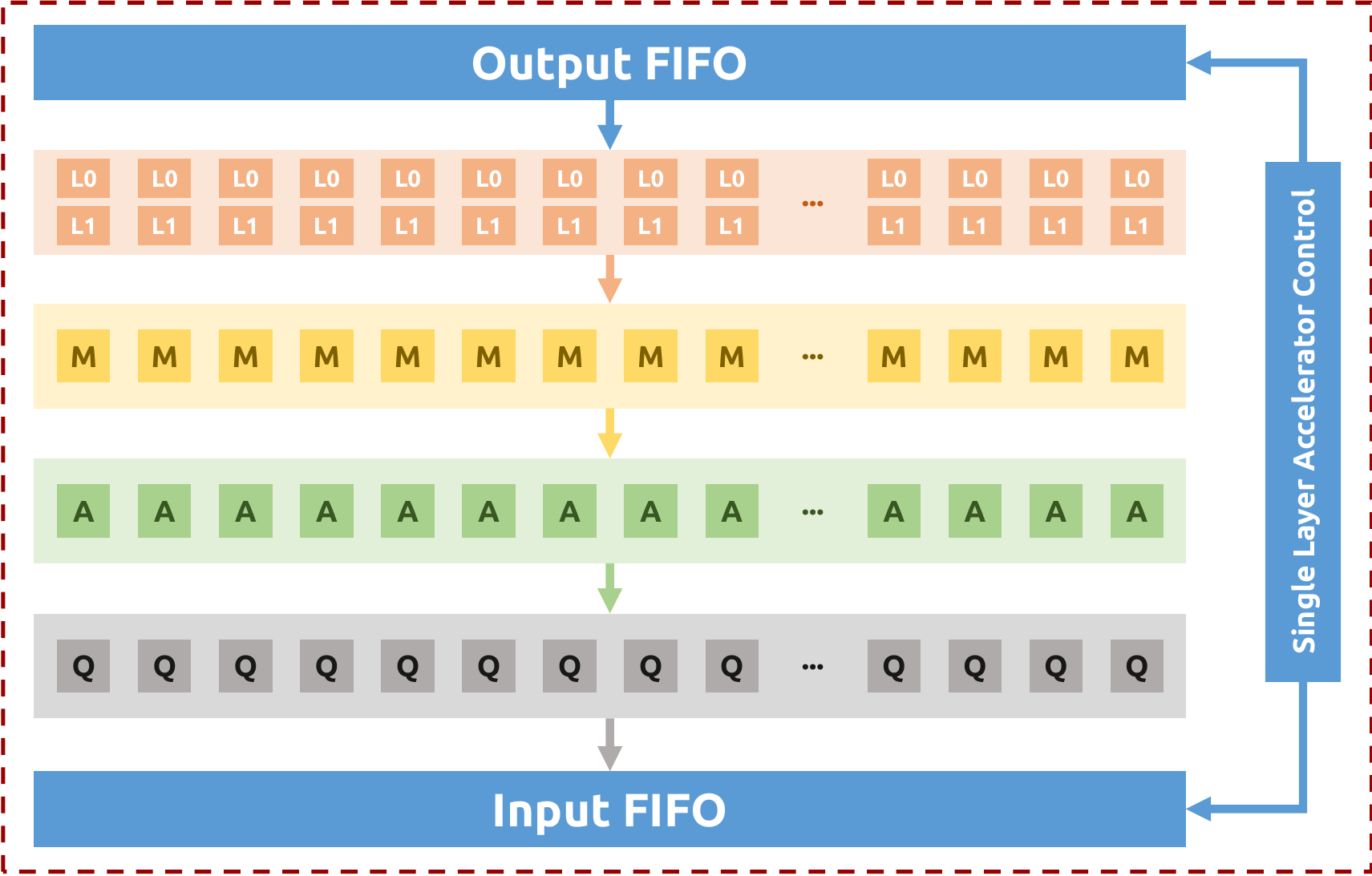}}
        \caption{\emph{BitSys}-based Single-Layer Accelerator}
        \label{bitsys_layer_structure}
    \end{minipage}
    \begin{minipage}{0.585\textwidth}
        \centerline{\includegraphics[width=\columnwidth]{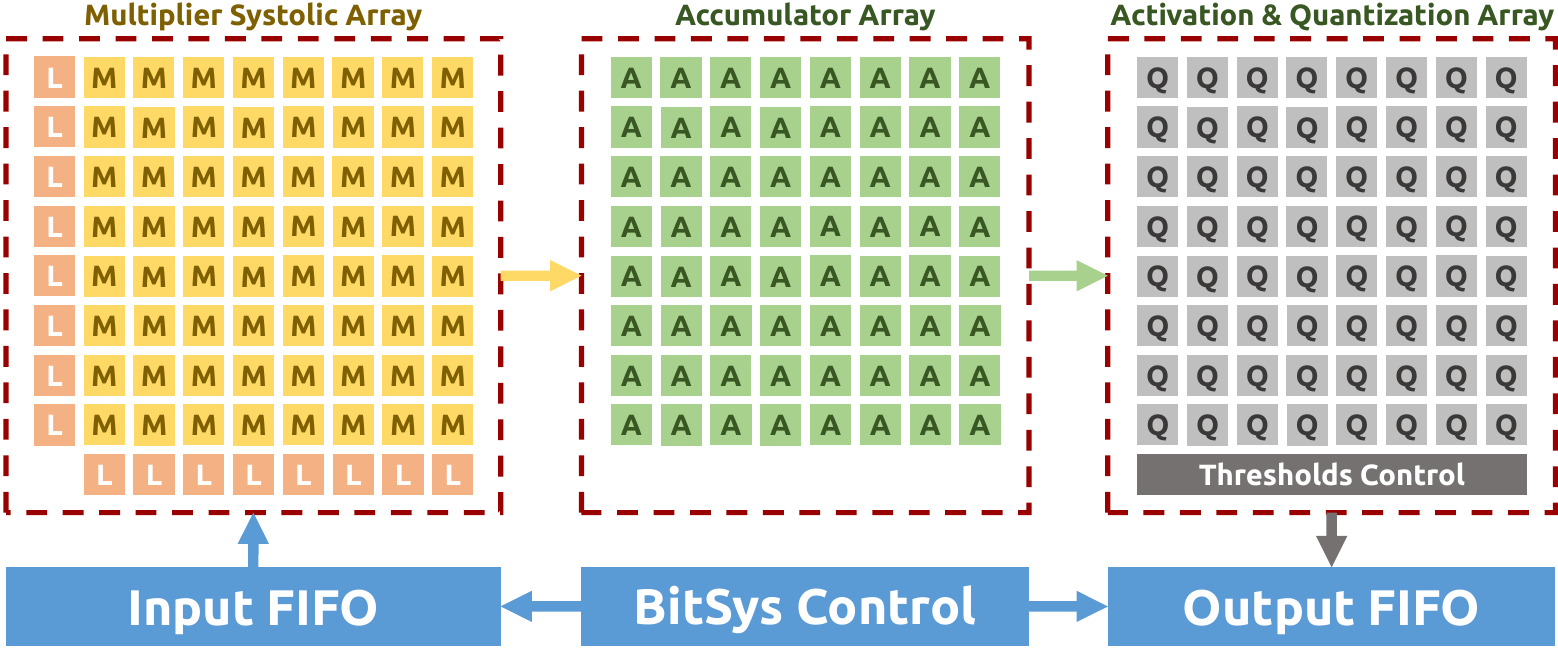}}
        \caption{\emph{BitSys}-based Systolic-Array Accelerator}
        \label{bitsys_systolic_structure}
    \end{minipage}
\end{figure*}

To evaluate the multiplier based on our \emph{BitSys} architecture, we implemented one single-layer accelerator and one systolic array accelerator as shown in~\autoref{bitsys_layer_structure} and~\autoref{bitsys_systolic_structure}. Both accelerators consist of four components: 1) Input Loader (Orange), 2) \emph{BitSys} Multiplier (Yellow), 3) Accumulator (Green), and 4) Activation Module (Gray). Both accelerators contain 64 multipliers. The single-layer accelerator implements these multipliers as 8 neurons. Each neuron consists of 8 multipliers and 16 input loaders. The systolic array accelerator implements these multipliers as an $8\times8$ systolic array with 16 input loaders. Both single-layer and systolic array accelerators implemented a state machine to control the inference of network models, which loads and stores the layer settings, like input length and precision, in a FIFO of FPGA. To reconfigure the multipliers for different layer precision, the state machine uses three clock cycles to load the precision data from FIFO and rewrite the registers for multiplier settings.

Considering the output of \emph{BitSys} multiplier is multi-channel, if we implement the corresponding accumulator and activation module for all channels, one multiplier needs to connect with eight accumulators and eight activation modules at maximum (for 1-bit mode). However, when the multiplier works on higher precision, the required accumulators and activation modules are less than 1-bit mode because of fewer output channels, leading to low hardware efficiency. Therefore, we connect each multiplier with one accumulator and activation module in both accelerators. To this end, we implemented a tree-structure-based pipelined input converter shown in~\autoref{accu_structure} for the accumulator to sum all channels of multiplier output: Multiplier outputs 16-bit data to this input converter as $in_{0-15}$. The left-shifters (Orange) in~\autoref{accu_structure} apply the bit weight, $2^i$, to $in_{i}$ by passing through four shifting-and-adding layers in this tree structure. Because if $A$ is a signed value, $A=-2^{n-1}a_{n-1}+\sum_{i=0}^{n-2} 2^{i}a_{i}$. $a_{i}$ is the bit value of A and $a_{n-1}$ is the sign bit. Therefore, we insert one value inverter (Neg. Block) in the first shifting-and-adding layer to negate the left-shifted sign bit. For different precision, different numbers of value inverters are enabled. For instance, for 8-bit dual-channel input, only the inverters connected with $in_{7}$ and $in_{15}$ are enabled. Furthermore, we applied the \emph{Multi-Thresholds} activation function from FINN~\cite{Umuroglu2017, Blott2018} to design our activation module, which fused the activation and output re-quantization as multi-thresholds. This activation function required 1/3/15/255 thresholds to generate 1/2/4/8-bit output. The number of thresholds smaller than the accumulator output is the final output. Therefore, to reduce resource consumption and improve hardware efficiency, we only implement one comparator in each activation module, sequentially loading the thresholds to compare with the accumulator output. 


%% file: arXiv/evaluation.tex
\subsection{Experiment Setup}

We evaluate the Multiplier (MUL), Multiply-Accumulator (MAC), and accelerator instances of our~\emph{BitSys} architecture on \emph{Ultra96-V2} FPGA platform (\emph{Zynq UltraScale+ ZU3EG}). Considering the discussion in~\autoref{background}-B, we selected the works of Liu et al.~\cite{Liu2023} as the baseline. All accelerators are evaluated by the TFC models we trained as the same as the network used in~\cite{Liu2023} with the \emph{Brevitas} and \emph{MNIST} dataset, which have been introduced in~\autoref{introduction}.

\begin{table*}[t]
    \centering
    \caption{Resource Consumption of Multipliers (MUL) and Multiply-Accumulators (MAC)}
    \resizebox{0.975\linewidth}{!}
    {
        \begin{tabular}{cccccccccccccccc}
            \toprule
            \multirow{2}{*}{Design} & \multicolumn{4}{c}{Instance Setting}                                                & \multicolumn{2}{c}{Resource Consumption} & \multirow{2}{*}{Frequency} & Total Delay & Area-Delay & Dynamic Power & PDP & \multicolumn{4}{c}{Compuation Cycles} \\
                                    & Type                  & Precision & Signed/Unsigned & Accurate/Approximate & LUT                 & FF                 & &   ($ns$)              & Products        &   ($mW$)     & ($mW\times ns$)        & BIN      & 2-bit      & 4-bit     & 8-bit     \\ \midrule
            MTree-base~\cite{Liu2023}              & \multirow{6}{*}{MUL}  & \multirow{6}{*}{1/2/4/8}   & \multirow{6}{*}{Both}            & \multirow{6}{*}{Accurate}             & 383                 & 42                 & 250$MHz$ & 3.820       & 1463.06       &87    &   332.34      & 1        & 1          & 1         & 1         \\
            MTree-pipe              &                       &     &              &               & 429                 & 282                & 375$MHz$ & 2.282         &978.98         &125   & 285.25     & 5        & 5          & 5         & 5         \\
            Bitshifter-base~\cite{Liu2023}         &                       &     &              &               & 345                 & 37                 & 300$MHz$ & 3.156      &1088.82         &107     & 337.69      & 1        & 1          & 1         & 1         \\
            Bitshifter-pipe         &                       &     &              &               & 337                 & 339                & 375$MHz$ & 2.171          & 731.627         &122    & 264.86   & 1        & 9          & 9         & 9         \\
            \textbf{BitSys-base}             &                       &     &              &               & 416                 & 463                & 500$MHz$ & 1.433        &596.128     &156       & 223.55      & 22       & 22         & 22        & 22        \\
            \textbf{BitSys-LUT}              &                       &     &              &               & 350                 & 525                & 500$MHz$ & 1.419         & 496.65     &159      & 225.62      & 22       & 22         & 22        & 22        \\ \midrule
            MTree-base~\cite{Liu2023}              & \multirow{6}{*}{MAC}       & \multirow{6}{*}{1/2/4/8}   & \multirow{6}{*}{Both}            & \multirow{6}{*}{Accurate}               & 398                 & 199                & 250$MHz$ & 3.397        &1352.01    &79       & 268.36       & 6        & 6          & 6         & 6         \\
            MTree-pipe              &                       &     &              &               & 495                 & 388                & 250$MHz$ & 2.828            & 1399.86      &102     & 288.46   & 10       & 10         & 10        & 10        \\
            Bitshifter-base~\cite{Liu2023}         &                       &     &              &               & 505                 & 198                & 300$MHz$ & 3.084          & 1425.27     &102      & 314.57     & 6        & 6          & 6         & 6         \\
            Bitshifter-pipe         &                       &     &              &               & 538                 & 506                & 375$MHz$ & 2.164          & 1164.23         &109    & 235.88   & 6        & 14         & 14        & 14        \\
            \textbf{BitSys-base}             &                       &     &              &               & 597                 & 633                & 375$MHz$ & 2.072          & 1236.98        &103    & 213.42   & 27       & 27         & 27        & 27        \\
            \textbf{BitSys-LUT}              &                       &     &              &               & 541                 & 689                & 500$MHz$ & 1.716          & 928.36          &134    & 229.94  & 27       & 27         & 27        & 27       \\ \bottomrule
        \end{tabular}
    }
    \label{mul_evaluation_tab}
\end{table*}
\begin{table*}[t]
    \centering
    \begin{minipage}{0.975\linewidth}
        \caption{Resource Consumption of Previous and BitSys Accelerators on \textit{Ultra96V2} FPGA Platform}
        \resizebox{1\columnwidth}{!}
        {
            \begin{tabular}{ccccccccccc}
                \toprule
                \multirow{2}{*}{Design}    & \multirow{2}{*}{Type}         & \multirow{2}{*}{Precision} & \multicolumn{2}{c}{LUT}                           & \multicolumn{2}{c}{FF}                            & \multicolumn{2}{c}{BRAM}                        & \multirow{2}{*}{Frequency} & \multirow{2}{*}{Latency/$\mu s$} \\
                                           &                               &                            & Number                 & Rate                     & Number                 & Rate                     & Number               & Rate                     &                            &                             \\ \midrule
                \multirow{2}{*}{Vivado IP~\cite{neda2022multi}} & \multirow{2}{*}{Single-Layer} & 8/8/8/8                    & \multirow{2}{*}{24090} & \multirow{2}{*}{34.14\%} & \multirow{2}{*}{22175} & \multirow{2}{*}{15.71\%} & \multirow{2}{*}{135} & \multirow{2}{*}{62.50\%} & \multirow{2}{*}{150$MHz$}    & 137.654                     \\
                                           &                               & 1/2/4/8                    &                        &                          &                        &                          &                      &                          &                            & 131.059                     \\ \midrule
                MTree - base~\cite{neda2022multi}               & Single-Layer                  & 1/2/4/8                    & 37020                  & 52.47\%                  & 22500                  & 15.94\%                  & 138                  & 63.89\%                  & 100$MHz$                     & 69.27                       \\
                Bitshifiter - base~\cite{neda2022multi}         & Single-Layer                  & 1/2/4/8                    & 42952                  & 60.87\%                  & 22486                  & 15.93\%                  & 138                  & 63.89\%                  & 125$MHz$                     & 56.658                      \\
                MTree - pipe               & Single-Layer                  & 1/2/4/8                    & 47163                  & 66.84\%                  & 42100                  & 29.83\%                  & 138                  & 63.89\%                  & 150$MHz$                     & 48.443                      \\
                Bitshifiter - pipe         & Single-Layer                  & 1/2/4/8                    & 50212                  & 71.16\%                  & 50393                  & 35.71\%                  & 138                  & 63.89\%                  & 150$MHz$                     & 48.799                      \\ \midrule
                \textbf{BitSys - LUT}               & Single-Layer                & 1/2/4/8                    & 46570                  & 66.00\%                  & 54352                  & 38.51\%                  & 138                & 63.89\%                  & 150$MHz$                     & 49.057                     \\
                \textbf{BitSys - LUT}               & Systolic Array                & 1/2/4/8                    & 44468                  & 63.02\%                  & 64176                  & 45.48\%                  & 139.5                & 64.58\%                  & 250$MHz$                     & 36.741                     \\ \bottomrule
            \end{tabular}
        }
        \label{bitsys_evaluation_tab}
    \end{minipage}
\end{table*}

\subsection{Multiplier and Multiply-Accumulator Comparison}

\autoref{mul_evaluation_tab} shows the implementation results from \emph{Vivado}: we implemented six instances, including one pure-Verilog-designed and one LUT-primitive-optimized variant, \emph{BitSys-base} and \emph{BitSys-LUT}, for 1/2/4/8-bit signed/unsigned MULs and MACs of our~\emph{BitSys}. As the baseline, we implement the MUL and MAC instances of \emph{Multiplier-Tree} and \emph{Bitshifter} from Liu et al.~\cite{Liu2023} as~\emph{MTee-base} and~\emph{Bitshifter-base}, supporting 1/2/4/8-bit signed/unsigned reconfigurable multiplication, and insert the registers between the sub-multipliers of \emph{Multiplier-Tree} and AND/Mask/Shifting stages of \emph{Bitshifter} as shown in~\autoref{related_works_fig}.2 and~\autoref{related_works_fig}.4 to create their pipelined instances,~\emph{MTee-pipe} and~\emph{Bitshifter-pipe}, for higher clock frequency. 

For MUL comparison, two \emph{Bitshifter} instances consume fewer LUTs with less total path delay than \emph{Multiplier-Tree} instances. The pipelined \emph{Multiplier-Tree} and \emph{Bitshifter} instances,~\emph{MTee-pipe} and~\emph{Bitshifter-pipe}, consume more LUTs and FFs than their basic instances,~\emph{MTee-base} and~\emph{Bitshifter-base}, with significant decrease in total path delay. Compared with these baseline instances, both MULs of \emph{BitSys} have the lower total path delay: the lowest total path delay belongs to \emph{BitSys-LUT}, which is $65.36\%$, $44.97\%$, $62.18\%$, and $33.51\%$ of~\emph{Bitshifter-pipe},~\emph{Bitshifter-base},~\emph{MTee-pipe}, and~\emph{MTee-base}. The LUT-primitive-optimization of \emph{BitSys-LUT} instances decreased the resource consumption and total path delay compared with \emph{BitSys-base}. The LUT consumption of \emph{BitSys-LUT} shows no advantages with the same or higher numbers than~\emph{MTee-base},~\emph{Bitshifter-base}, and~\emph{Bitshifter-pipe} as $100.00\%$, $101.45\%$, and $103.86\%$.However, we discussed the \emph{Area Delay Products} (ADPs) in \autoref{mul_evaluation_tab}, which are the products between LUT consumption and total path delay. The lowest ADP of \emph{BitSys} instances implies that our work achieved an efficient design with a good balance between performance and resource utilization. Based on the post-implementation timing simulation in \emph{Vivado}, we analyzed the power consumption of all MUL instances with 16000 times random multiplication under the highest available clock shown in~\autoref{mul_evaluation_tab}. Our~\emph{BitSys} instances have the highest power consumption. However, their \emph{Power Delay Products} (PDPs) are lower than the other four instances, which are the products between power and total path delay. This means that our design has better power efficiency and achieves a good balance between minimizing power usage and maximizing speed. Moreover, the \emph{Computation Cycles} column in~\autoref{mul_evaluation_tab} shows that \emph{BitSys} architecture has a longer pipeline path in computation than other instances, which explains the low total path latency and high FF consumption of our work. Differing from the MUL instances, MAC instances of \emph{Multiplier-Tree} cost less LUT than \emph{Bitshifter} because we fused the accumulator input converter design of \emph{Multiplier-Tree} by summing the results of sub-multipliers and passing it to higher precision multipliers. The output of this multiplier is the sum of all channels. Following the same trend as MULs, The MAC instances of our \emph{BitSys} consume more resources and power than other instances with less total path delay, lower ADP, and lower PDP. The low power consumption of MACs compared with MULs is caused by the different testbench and longer pipeline as shown in \emph{Computation Cycles} column. We simulated the MACs with 4096 times random multiplication and accumulation for each precision. Before starting the next round of computation for another precision, MACs need to wait to finish the accumulation of the current precision. In summary, our~\emph{BitSys} architecture has a better design optimization between the balance of hardware consumption, power usage, and processing speed, supporting the highest clock frequency with the lowest total path delay.

\subsection{Neural Network Accelerator Comparison}

\autoref{bitsys_evaluation_tab} is the implementation and real measurement results of accelerators we implemented on \emph{Ultra96-V2} platform, including six single-layer accelerator based on~\emph{Vivado IP},~\emph{MTee-base},~\emph{MTee-pipe},~\emph{Bitshifter-base},~\emph{Bitshifter-pipe}, and~\emph{BitSys-LUT}, and one systolic array accelerator based on~\emph{BitSys-LUT} to compare the difference between single-layer accelerator architecture and systolic array architecture. The column of \emph{Latency/$\mu s$} is the single frame inference delay by averaging the total inference latency of 1000 MNIST inputs.

\emph{Vivado IP}-based single-layer accelerator consumes the least LUTs and FFs with the longest single-frame inference latency compared with other accelerators because it does not support multi-precision multiplication. Both~\emph{Bitshifter} accelerators cost more hardware resources than \emph{Multiplier-Tree}. The pipelined accelerator of both \emph{Multiplier-Tree} and \emph{Bitshifter} support higher clock frequency than their basic accelerators. According to the total path delay of MUL and MAC shown in~\autoref{mul_evaluation_tab}, in principle, the single-layer accelerators of~\emph{BitSys-LUT} and~\emph{Bitshifter-pipe} should support higher clock frequency than~\emph{MTee-base},~\emph{MTee-pipe}, and~\emph{Bitshifter-base}. However, because the single-layer accelerator contains one more complex state machine than the systolic array for data streaming control to load the activations and weights from DDR to neurons and schedule the computation with a limited number of neurons and multipliers, 150$MHz$ is the highest frequency that can be supported in our current single-layer accelerator architecture. The systolic array accelerator implemented with~\emph{BitSys-LUT} consumes $95.49\%$ of LUTs but $118.07\%$ of FFs with 250$MHz$ compared with its single-layer accelerator. Comparing the two structures shown in~\autoref{bitsys_layer_structure} and~\autoref{bitsys_systolic_structure}, systolic array accelerator of~\emph{BitSys-LUT} requires much fewer input loaders that single-layer accelerator, which causes the LUT consumption decreasing shown in~\autoref{bitsys_evaluation_tab}. For the average inference latency, all accelerators implemented with multi-channel multi-precision multiplier have a high speed-up compared with \emph{Vivado IP}-based accelerator. The single-layer accelerators of~\emph{Bitshifter-pipe} and~\emph{MTee-pipe} are $116.1\%$ and $142.99\%$ faster than~\emph{Bitshifter-base} and~\emph{MTee-base} because of higher clock frequency. Compared with the single-layer accelerator of~\emph{Bitshifter-pipe} and~\emph{MTee-pipe}, and~\emph{BitSys-LUT},~\emph{BitSys-LUT} instance is $0.53\%$ and $1.25\%$ slower than~\emph{Bitshifter-pipe} and~\emph{MTee-pipe} with same frequency. Considering the \emph{Computation Cycles} shown in~\autoref{mul_evaluation_tab},~\emph{BitSys-LUT} instance infers slower because of its long pipeline path. For the same reason, the single-layer accelerator of~\emph{Bitshifter-pipe} is $0.73\%$ slower than~\emph{MTee-pipe}. However, because the systolic array structure simplified the data streaming control in a single-layer accelerator, the inference latency of~\emph{BitSys-LUT} benefits both from the higher clock frequency and denser computation, which can highly efficiently utilize the fully pipelined design in our~\emph{BitSys} architecture. Therefore, the systolic array accelerator of~\emph{BitSys-LUT} supports 250$MHz$ and is $356.71\%$, $188.54\%$, $148.77\%$, $131.85\%$, $132.82\%$, and $133.52\%$ faster than the single-layer accelerators of~\emph{Vivado IP},~\emph{MTee-base},~\emph{MTee-pipe},~\emph{Bitshifter-base},~\emph{Bitshifter-pipe}, and~\emph{BitSys-LUT} with mixed-precision TFC network.

%% file: arXiv/conclusion.tex
In this manuscript, we present one multiplier design based on fully pipelined bitwise systolic array architecture, \emph{BitSys}, supporting the runtime reconfigurable multi-precision multi-channel multiplication. The evaluation shows that our \emph{BitSys} architecture has a low critical path delay to support higher clock frequency compared with previous works. In the acceleration of the mixed-precision network model, our work is more than $131.85\%$ faster than original \emph{Multiplier-Tree} and \emph{Bitshifter} architecture and about $356.71\%$ faster compared with Vivado-IP-based accelerator. For our future work, we plan to explore the ASIC implementation of our~\emph {BitSys} architecture with emerging memory technologies, such as \emph{Racetrack Memory} (RTM). 

%% file: arXiv/acknowledgements.tex
This work is supported by the~\emph{Center for Scalable Data Analytics and Artificial Intelligence} (ScaDS.AI)~\emph{Dresden/Leipzig} and~\emph{Deutsche Forschungsgemeinschaft} (DFG) under the~\emph{X-ReAp} project (Project number 380524764).